\documentclass[aps,prb,twocolumn,superscriptaddress]{revtex4}

\usepackage{epsfig}
\usepackage{graphicx}
\usepackage{dcolumn}
\usepackage{bm}

\def\vec#1{\mathbf{#1}}
\def\figfont{}

\begin{document}

\input epsf.sty

\newcommand{\lm}{\lambda_{\mathrm{max}}}
\newcommand{\lav}{\lambda_{\mathrm{av}}}
\newcommand{\ks}{h_{\mathrm{KS}}}
\newcommand{\tr}{\hbox{Tr}}

\title{Chaotic properties of spin lattices near second-order phase transitions}
\author{A. S. de Wijn}
\email{astrid@dewijn.eu, dewijn@fysik.su.se}
\affiliation{Department of Physics, Stockholm University, 106 91 Stockholm, Sweden}
\author{B. Hess}
\affiliation{Institute for Theoretical Physics, University of Heidelberg, Philosophenweg 19, 69120 Heidelberg, Germany}
\author{B. V. Fine}
\email{B.Fine@thphys.uni-heidelberg.de}
\affiliation{Institute for Theoretical Physics, University of Heidelberg, Philosophenweg 19, 69120 Heidelberg, Germany}
\affiliation{Department of Physics, School of Science and Technology, Nazarbayev University 53 Kabanbai Batyr Ave., Astana 010000, Kazakhstan }
\affiliation{Skolkovo Institute of Science and Technology, 100 Novaya Str., Skolkovo, Moscow Region 143025, Russia  }

\begin{abstract}
We perform a numerical investigation of the Lyapunov spectra of chaotic dynamics in lattices of classical spins in the vicinity of second-order ferromagnetic and antiferromagnetic phase transitions. On the basis of this investigation, we identify a characteristic of the shape of the Lyapunov spectra, the ``G-index", which exhibits a sharp peak as a function of temperature at the phase transition, provided the order parameter is capable of sufficiently strong dynamic fluctuations.
As a part of this work, we also propose a general numerical algorithm for determining the temperature in many-particle systems, where kinetic energy is not defined.
\end{abstract}

\maketitle 

\section{Introduction}
\label{Intro}

The notion of chaos is often invoked in statistical physics to justify the ergodicity assumption.  However, the relation
between the primary characteristics of chaos, namely the Lyapunov exponents, and the equilibrium properties of
many-particle systems still remains elusive. A particularly interesting issue in this regard is the temperature        
dependence of Lyapunov exponents near phase transitions. Lyapunov exponents characterize the sensitivity of phase-space    
trajectories with respect to small deviations of initial conditions, while
phase transitions are often accompanied by strong fluctuations and the appearance of long-range correlations. Because of
the striking nature of phase transitions, one may
wonder if such dramatic changes involve chaos as well and if there are universal features to be found in the Lyapunov   
exponents close to phase transitions.

Most of the relevant investigations so far  have been limited to the largest Lyapunov exponents    
and often reported dramatic signatures of phase transitions in their temperature
dependences~\cite{Butera-87,Mehra-97,KwonPark1997,PhysRevLett.79.4361,Caiani-98,Caiani-98A,PhysRevE.57.6599,PhysRevLett.80.692,RivistaCasetti,0295-5075-55-2-164,PhysRevE.61.5171,Casetti-00,poschlj}. It should be noted, however, that, for some systems, these signatures likely originate from the infinite range of particle-particle                 
interactions~\cite{PhysRevLett.80.692,PhysRevE.57.6599}, while, for others, as we explain later, they are not intrinsic
to Lyapunov exponents but rather reflect the nonanalytic behavior of the temperature with respect to the total energy near a phase transition and would disappear if Lyapunov exponents are plotted as functions of the total energy.  At the same time, the investigations of Refs.\cite{PhysRevLett.79.4361,Caiani-98,Caiani-98A} (reviewed in Ref.\cite{Casetti-00} )  indicated that a quantity closely related to Lyapunov exponents, namely, the curvature of the configuration space in the geometrical formulation of the dynamics, exhibits sharply increasing fluctuations at phase transitions.                 

In general, Hamiltonian dynamics in an $N$-dimensional phase space generates not one but $N$ Lyapunov exponents         
organized in pairs of equal absolute values and opposite signs.                                                        
The entire Lyapunov spectra have been investigated so far only across first-order phase                                 
transitions~\cite{dellagoposchphasetransspectrum1996,poschlj2,Bosetti-14}.
In this paper, we present a detailed investigation of Lyapunov spectra as a function of temperature for lattices of classical spins with nearest-neighbor interaction in the vicinity of ferromagnetic (FM) and antiferromagnetic (AF) second-order phase transitions.
We introduce a characteristic of the shape of the Lyapunov spectra, namely the ``$G$-index'', which exhibits a peak at the phase transition both as a function of temperature {\it and energy}, provided the order parameter is capable of sufficiently strong dynamic fluctuations.
The present work builds on our earlier investigations of Lyapunov instabilities in classical spin lattices at infinite temperature~\cite{chaosspins,chaosspins2}, where, in particular, we showed that the lattices are all chaotic with the exception of the Ising case.

\section{General formulation and numerical aspects}

We consider cubic lattices of $N_s$ classical spins with periodic boundary conditions and the nearest-neighbor
interaction Hamiltonian:
\begin{equation}
{\mathcal H} =\sum_{i,j(i),i<j} J_x S_{ix} S_{jx}+J_y S_{iy} S_{jy}+J_z S_{iz} S_{jz}~,
\label{H}
\end{equation}
where $(S_{ix}, S_{iy}, S_{iz}) \equiv {\mathbf S}_i$ are the three projections of the $i$th classical spin normalized by the condition ${\mathbf S}_i^2 = 1$, and $J_x, J_y$ and $ J_z$ are the coupling constants, which we choose such that $J_x^2 + J_y^2 + J_z^2 = 1$. The notation $j(i)$ indicates the nearest neighbors of the $i$-th spin.
Various alternatives to the Hamiltonian in Eq.~(\ref{H}) will be discussed in Sec.~\ref{sec:alternativedynamics}.

Our procedure for computing the spectrum of Lyapunov exponents $\{ \lambda_i \}$ is described in Ref.~\cite{chaosspins2}.  It follows the standard approach of Ref.~\cite{Benettin-80}. Index $i$ in the above notation orders the Lyapunov exponents in decreasing order, with $\lambda_1 \equiv \lm$ being the largest positive Lyapunov exponent.  Due to the demanding nature of the numerical calculations of the full Lyapunov spectrum (order $N_s^2$ per time step combined with long convergence times for small exponents), we have had to restrict ourselves to lattices of $8\times 6\times 4$. For this system size, the finite-size effects on the Lyapunov exponents are already small~\cite{chaosspins2} (for more details, see Appendix~\ref{appendix:finitesize}). 

We numerically integrate the equations of motion associated with the Hamiltonian~(\ref{H}),
$\dot{\mathbf S}_i = {\mathbf S}_i \times {\mathbf h}_i$,
where
$\mathbf h_i = \sum_{j(i)} J_x S_{jx} {\mathbf e}_x + J_y S_{jy} {\mathbf e}_y + J_z S_{jz} {\mathbf e}_z$
is the local field.
Here  ${\mathbf e}_x$, ${\mathbf e}_y$ and ${\mathbf e}_z$ are orthogonal unit vectors. We use a fourth-order Runge-Kutta algorithm with time step 0.005.
During the time of our simulations, typically equal to 20000, the total energy is conserved with absolute accuracy better than $10^{-6}$.
The initial conditions corresponding to a given value of the total energy of the system are selected using the routine described in Ref.~\cite{chaosspins2}, which draws them from a uniform distribution of the energy shell.

\section{Determining temperature and identifying phase transitions}
\label{Temperature}

The total energy $E$ determines the temperature $T$ of the system. However, this temperature cannot be found using the average kinetic energy per particle, because the spin Hamiltonian cannot be decomposed into a quadratic-in-momentum kinetic energy and a momentum-independent potential energy~\cite{Ben-diploma}. Below we describe a more general algorithm applicable to any system with smooth dynamics and short-range interactions. (A system-specific alternative to this algorithm would be to simulate a thermal bath or to extract temperature from the correlations between spin polarizations and local fields.)

Our algorithm is based on the definition 
$1/T \equiv d S/d E$, where $S$ is the entropy of the system, which, is, in turn, defined (after setting $k_\mathrm{B} = 1$) as $S\equiv \ln V(E)$.  Here  $V(E)$ is the $(N-1)$-dimensional volume of the energy shell in the $N$-dimensional many-particle phase space. The above definitions lead to
\begin{equation}
\frac{1}{T} = \frac{1}{V(E)} \frac{d V(E)}{d E},
\label{T}
\end{equation}
which implies that the volume of the energy shell changes nearly exponentially as a function of energy with the characteristic constant equal to the inverse temperature.
For this reason, obtaining the above constant by random Monte-Carlo sampling of the entire many-particle phase space is not feasible. Instead, our algorithm consists of the following three steps: (i) It locates one point on any given energy shell using a dissipative dynamics routine introduced in Ref.~\cite{chaosspins2}. (ii) It randomly samples that energy shell using sequential {\it energy-conserving} rotations of randomly chosen spins around the directions of their local fields by random angles. (iii) Finally, it explores the vicinity of each thus obtained point on the energy shell by tiny {\it energy non-conserving} rotations of each spin around a randomly chosen axis perpendicular to spin's direction. The small angles for these rotations are drawn from a Gaussian distribution around zero.  (For the $8\times6\times4$ lattices considered, we used a standard deviation of $0.014$~rad.) Since $V(E)$ grows exponentially with energy, an energy increase as a result of step (iii) is more likely than an energy decrease.
We recover the value of temperature by first obtaining the mean and the mean-squared changes of energy, $\langle \Delta E\rangle$ and $\langle \Delta E^2 \rangle$, respectively, and then substituting them into the formula
\begin{equation}
{T} = \frac{\langle \Delta E^2 \rangle }{2 \langle \Delta E\rangle}~,
\label{T-alg}
\end{equation}
which is derived in Appendix~\ref{T-appendix}. 

\begin{figure}
\epsfig{figure=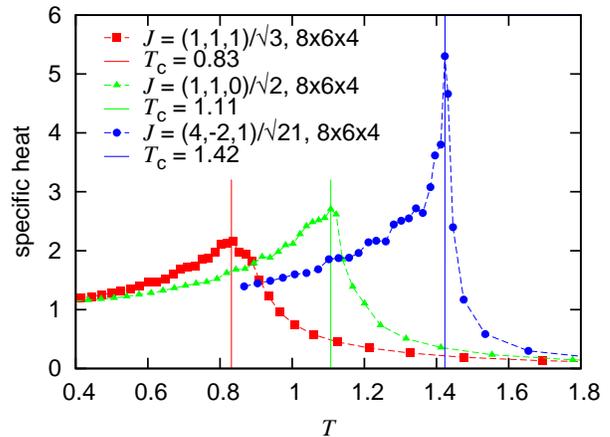,angle=270,width=8.6cm}
\caption{
Temperature dependence of the specific heat for three different lattices in the vicinity of their respective phase transitions.  We identify the phase transition as the point of maximum specific heat.
\label{fig:CCC}}
\end{figure}

After obtaining $E(T)$, we find the specific heat as $N C(T) = dE/dT$. In the thermodynamic limit, $C(T)$ exhibits lambda-point singularity at the FM and AF phase transitions. Since this singularity is washed out by the finite-size effects, we identify the phase-transition temperature $T_c$ with the maximum of $C(T)$ --- see figure~\ref{fig:CCC}.  In particular, for an $8 \times 6 \times 4$ lattice with the Heisenberg Hamiltonian, we thereby obtain $T_c=0.83$, which is the same as the appropriately rounded thermodynamic value that can be extracted from Refs.~\cite{Deng-05,Brown-06}. The size dependence of the specific heat is illustrated in Appendix~\ref{appendix:finitesize}. 

In our systems, $E=0$ corresponds to infinite temperature, while $E < 0$ and $E>0$ correspond to positive and negative temperatures respectively. Cubic spin lattices with nearest-neighbor interactions are bipartite, in the sense that they can be divided into two sublattices such that spins of one sublattice interact only  with the spins of the other sublattice. The reversal of all spin coordinates for one sublattice changes the sign of $E$ while leaving the volume of the corresponding phase space elements the same. As a result, the volumes of energy shells $V(E)$ are symmetric with respect to $E=0$, i.e. $V(E) = V(-E)$. This symmetry implies that, if an AF transition occurs at temperature~$T_c$, then, in the same system, an FM transition occurs at temperature~$-T_c$. We define the order parameters as 
$\phi  = \left| \frac{1}{N_s} \sum_i (\pm 1) \vec{S}_i\right|$,
where the FM order implies all signs $+1$, while the AF order implies  +1 and -1 alternating between adjacent lattice sites.

 Despite the above symmetry of $V(E)$, the Lyapunov spectra are, in general, not symmetric with respect to $E=0$, because the reversal of all three projections of a spin does not preserve their Poisson brackets (see Ref.~\cite{chaosspins2})  and hence changes the character of the dynamics. The only symmetric case is the $XX$-interaction characterized by $J_x = J_y \neq 0$ and $J_z = 0$. In this case, one can reverse only $x$- and $y$-components for the spins of one of the two sublattices without reversing their $z$-components, thereby protecting the Poisson brackets and, at the same time, reversing the energy. We further note, that, as illustrated in Appendix~\ref{reversal}, the Lyapunov spectra of bipartite lattices do not change under the simultaneous sign reversal of energy $E$ and the sign of one of the three coupling constants.

\begin{figure}
\medskip\vskip2.5ex
\hskip1.5cm{\figfont (a)}\hfill\strut\\[-8.5ex]
\epsfig{figure=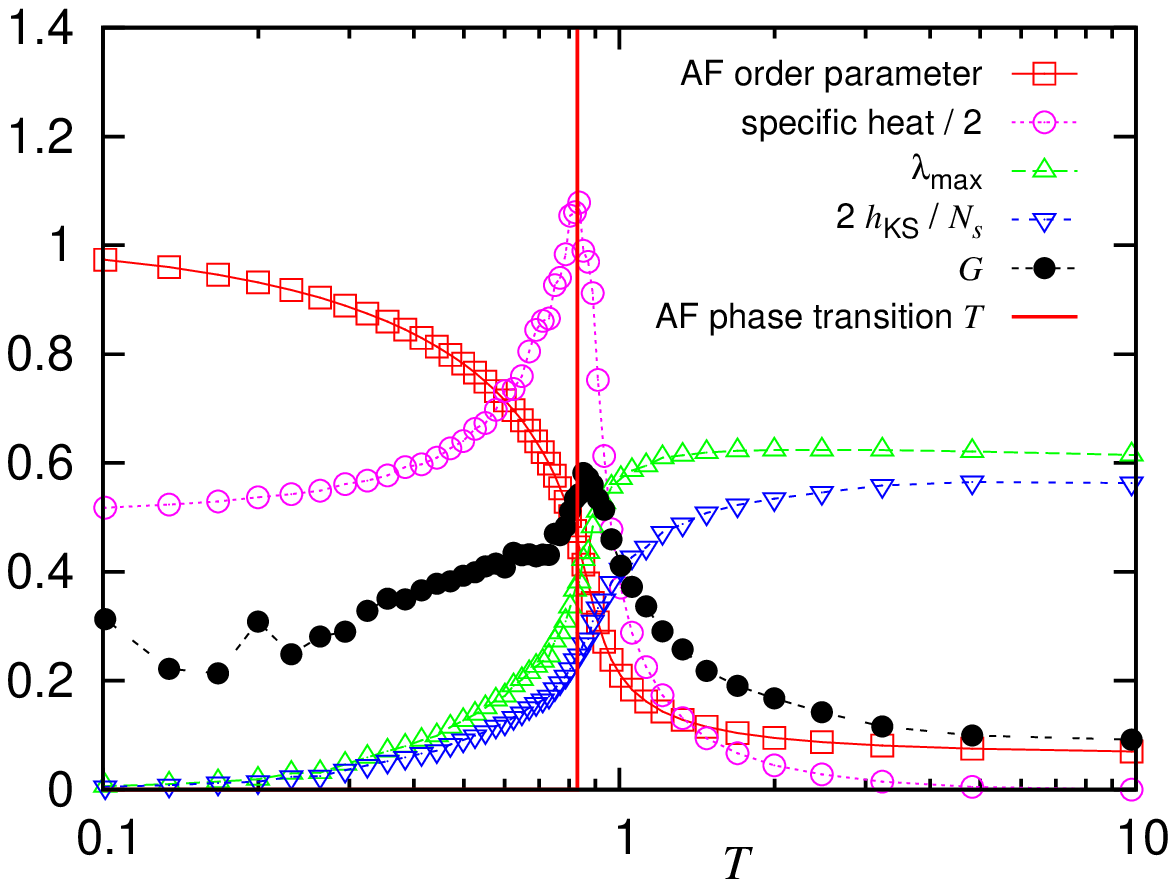,angle=270,width=8.6cm}
\vskip-1mm
\medskip\vskip2.5ex
\hskip1.5cm{\figfont (b)}\hfill\strut\\[-8.5ex]
\epsfig{figure=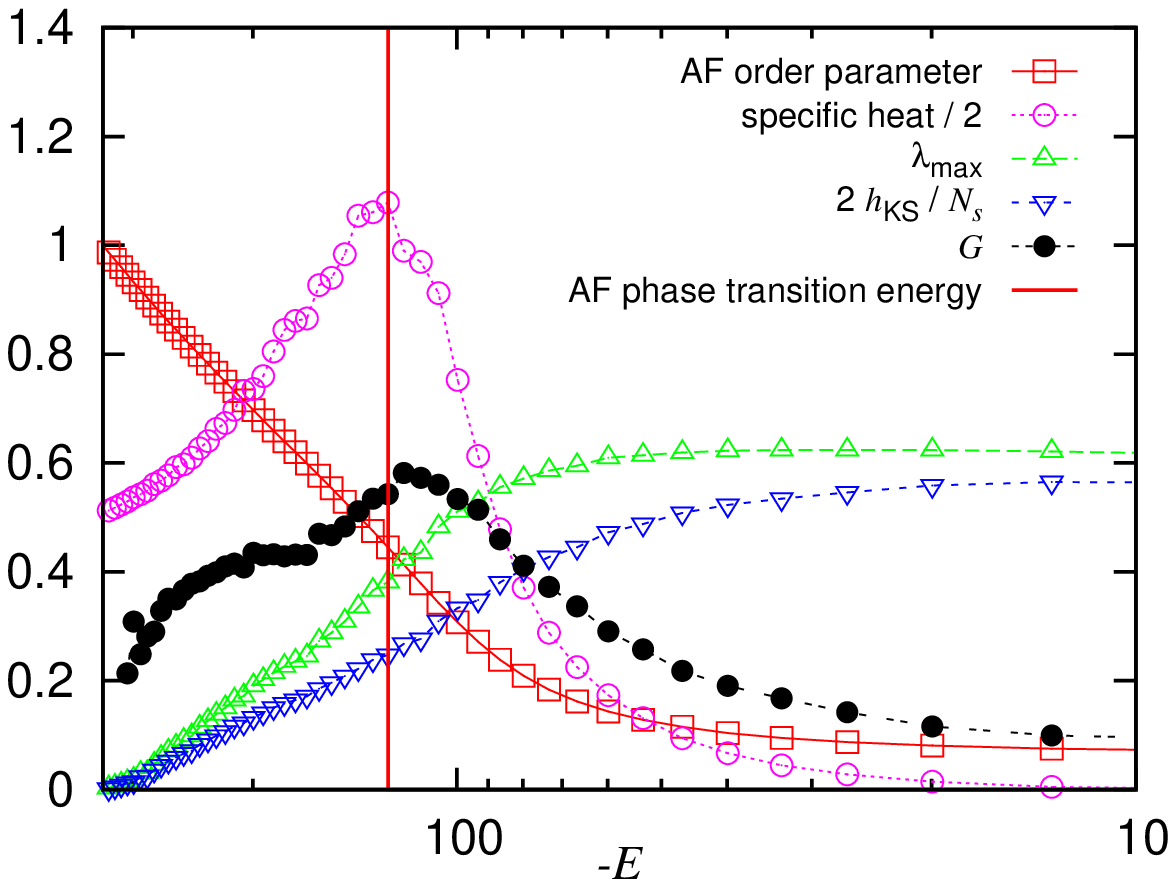,angle=270,width=8.6cm}
\vskip-1mm
\caption{
\label{fig:heisenberg}
Specific heat, AF order parameter, the largest Lyapunov exponent, Kolmogorov-Sinai entropy and the $G$-index as functions of (a) temperature and (b) energy  for the $8\times 6 \times 4$ lattice with Heisenberg interaction $J_x = J_y = J_z = 1/\sqrt{3}$. The positions of the AF phase transition are indicated by the vertical lines.
}
\end{figure}

\begin{figure}
\epsfig{figure=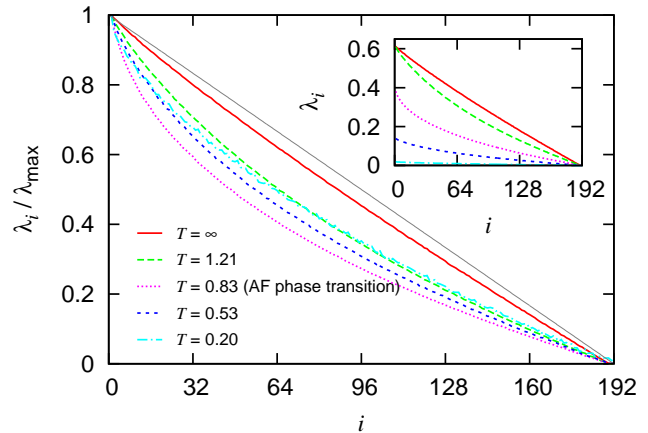,angle=270,width=8.6cm}
\caption{
\label{fig:spectra}
Examples of Lyapunov spectra for the lattice used for Fig.~\ref{fig:heisenberg}.
The thin gray straight line extending along the diagonal of the main plot is drawn to illustrate the geometrical interpretation of the $G$-index given in the text.
}
\end{figure}

\section{Numerical results: G-index}
\label{Results}

We now turn to the results of our simulations for the case of the Heisenberg interaction  $J_x = J_y = J_z = 1/\sqrt{3}$ at positive temperatures. Fig.~\ref{fig:heisenberg} shows the energy and temperature dependences of $\lm$ and the Kolmogorov-Sinai entropy $\ks$ (equal to the sum of all positive Lyapunov exponents) together with the specific heat and the order parameter.   Several examples of complete Lyapunov spectra are presented in Fig.~\ref{fig:spectra}.

Comparing Figs.~\ref{fig:heisenberg}~(a) and~(b), we observe that both $\lm$ and $\ks$ exhibit a steep change across the AF phase transition as functions of temperature but not as functions of energy.
This behavior does not change with system size (see Appendix~\ref{appendix:finitesize}).
In general, such behavior is expected for any smooth function of energy $f(E)$, which is then converted to a function of temperature $\tilde{f}(T) \equiv f(E(T))$. For the latter function, $d\tilde{f}/dT = {df \over dE} {dE \over dT} = {df \over dE} C(T)$. Since $C(T)$ exhibits a singularity at the phase transition, so does $d\tilde{f}/dT$.
In other words, the steep changes of $\lm(T)$ and $\ks(T)$ around $T=T_c$ as such indicate only the change of the energy-temperature relation rather than an intrinsic sensitivity of Lyapunov instabilities to phase transitions.

The examples of spectra shown in Fig.~\ref{fig:spectra}, nevertheless indicate that the phase transition influences the shape of the Lyapunov spectra: the closer the temperature to $T_c$, the more curved the spectrum. We quantify this shape change by a simple ratio, which we call the ``$G$-index": 
\begin{equation}
G = \frac{N \lm}{2 \ks}-1~.
\label{G}
\end{equation}
It represents the ratio of the total area between the spectrum and the diagonal line extending in Fig.~\ref{fig:spectra} from $(0,1)$ to $(N_s,0)$, divided by the area under the spectrum.
The $G$-index is plotted in Fig.~\ref{fig:heisenberg}.  It exhibits a sharp peak at the phase transition as a function of temperature and also a clear maximum at the corresponding energy. The size dependence of $G(T)$ is illustrated in Appendix~\ref{appendix:finitesize}.

\begin{figure*}
\parbox[t]{6.3cm}{\hskip9mm{\figfont (a)}\hfill\strut\\[-10mm]
    \epsfig{figure=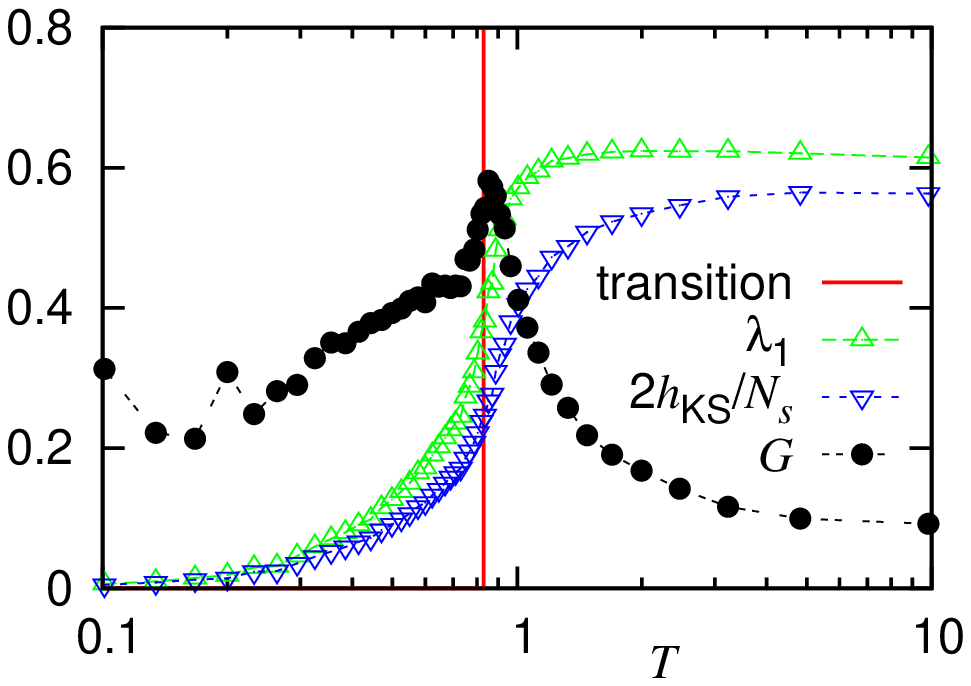,angle=270,width=6.3cm}
}
\parbox[t]{6.3cm}{\hskip9mm{\figfont (b)}\hfill\strut\\[-10mm]
    \epsfig{figure=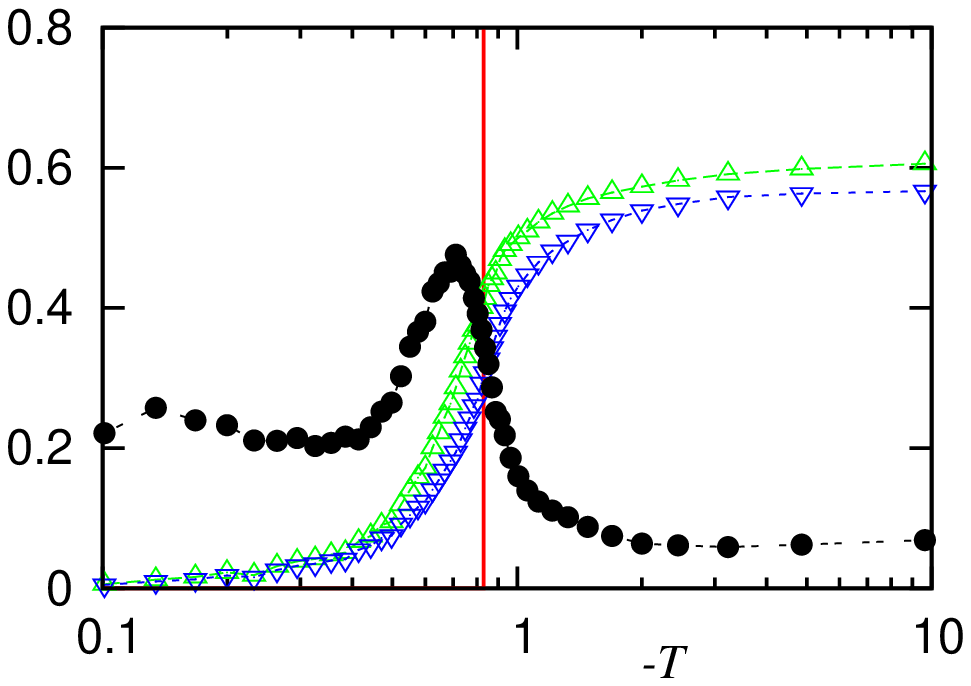,angle=270,width=6.3cm}
}\\[2.5ex]
\parbox[t]{6.3cm}{\hskip9mm{\figfont (c)}\hfill\strut\\[-10mm]
    \epsfig{figure=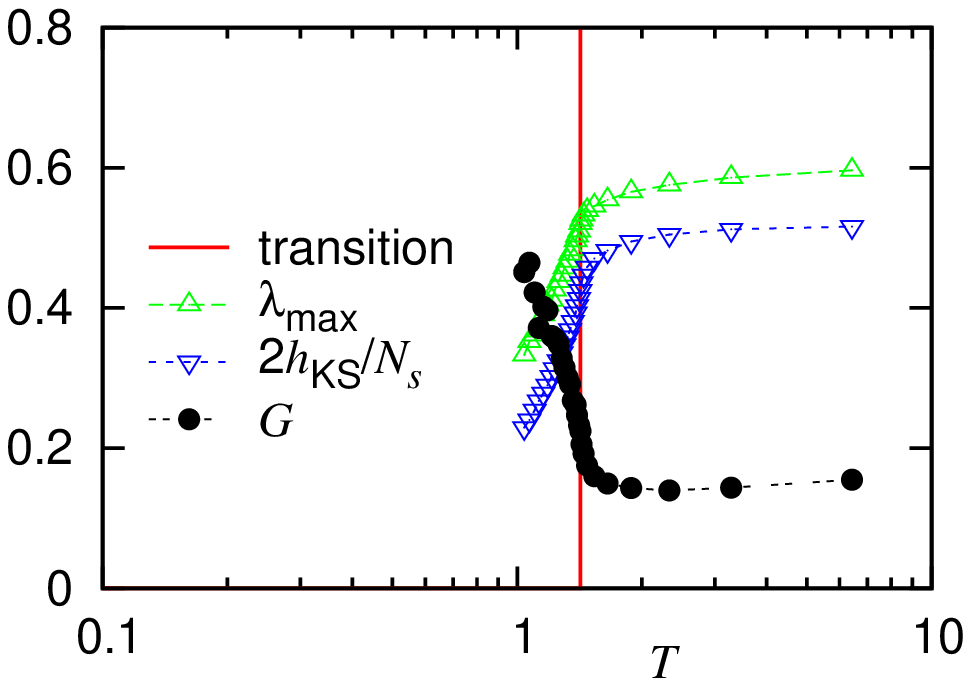,angle=270,width=6.3cm}
}
\parbox[t]{6.3cm}{\hskip9mm{\figfont (d)}\hfill\strut\\[-10mm]
    \epsfig{figure=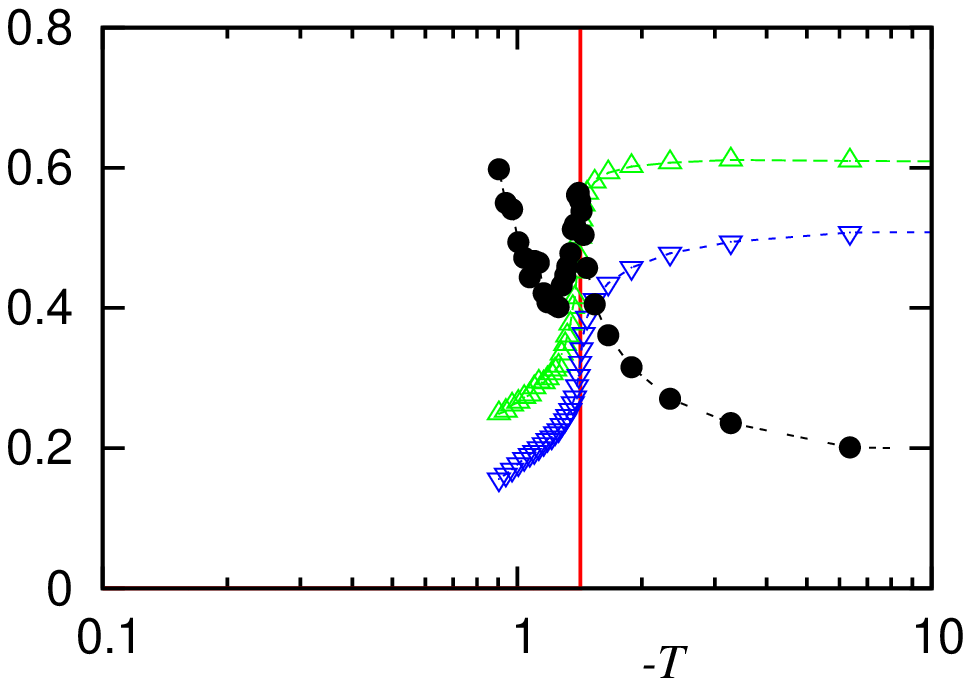,angle=270,width=6.3cm}
}\\[2.5ex]
\parbox[t]{6.3cm}{\hskip9mm{\figfont (e)}\hfill\strut\\[-10mm]
    \epsfig{figure=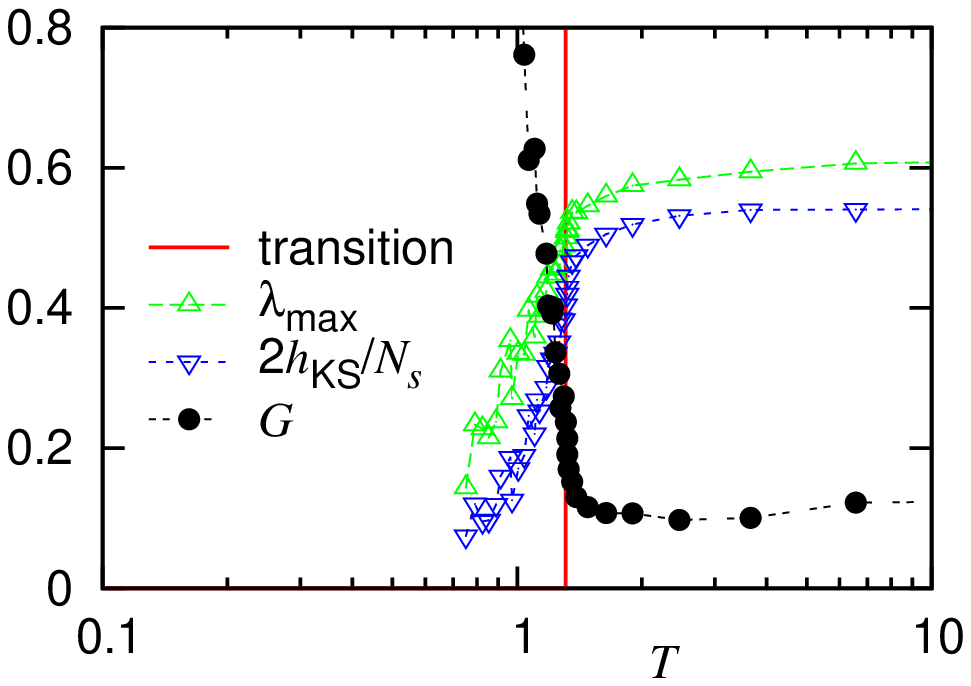,angle=270,width=6.3cm}
}
\parbox[t]{6.3cm}{\hskip9mm{\figfont (f)}\hfill\strut\\[-10mm]
    \epsfig{figure=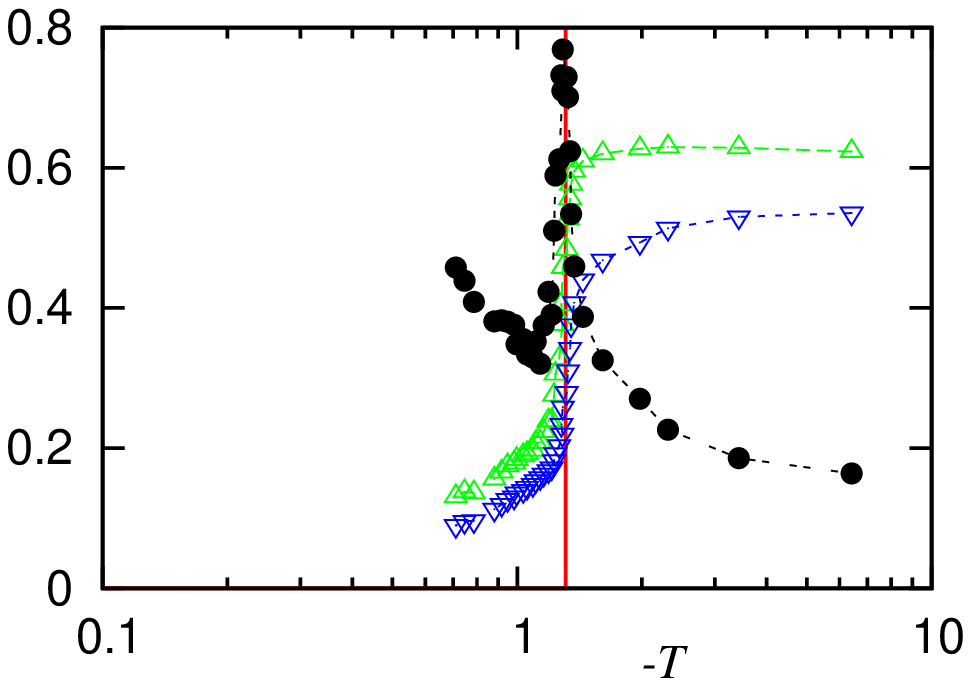,angle=270,width=6.3cm}
}\\[2.5ex]
\parbox[t]{6.3cm}{\hskip9mm{\figfont (g)}\hfill\strut\\[-10mm]
    \epsfig{figure=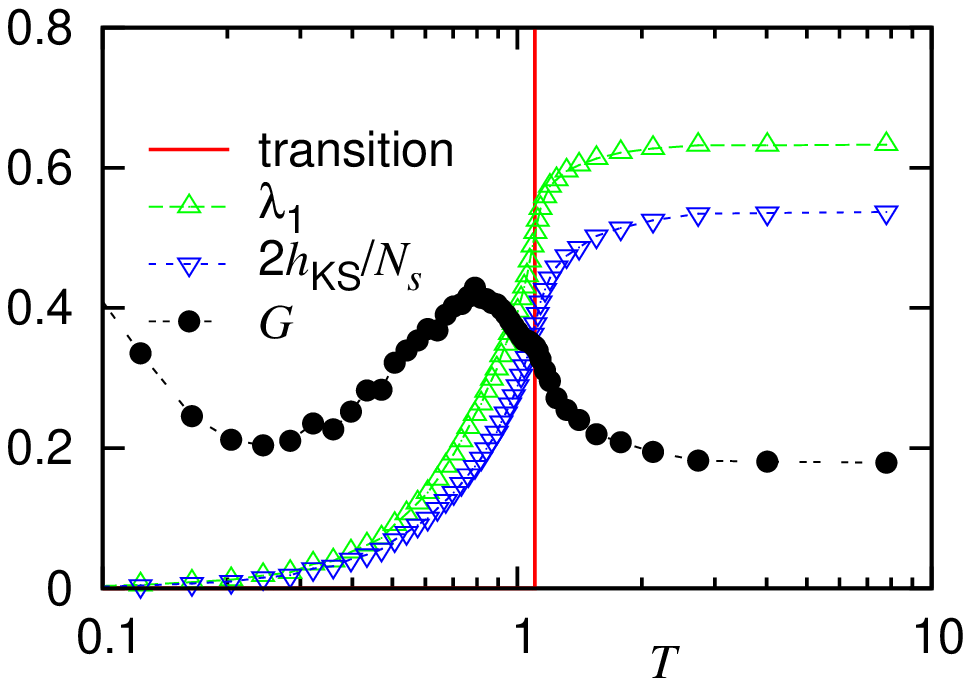,angle=270,width=6.3cm}
}
\parbox[t]{6.3cm}{\hskip9mm{\figfont (h)}\hfill\strut\\[-10mm]
    \epsfig{figure=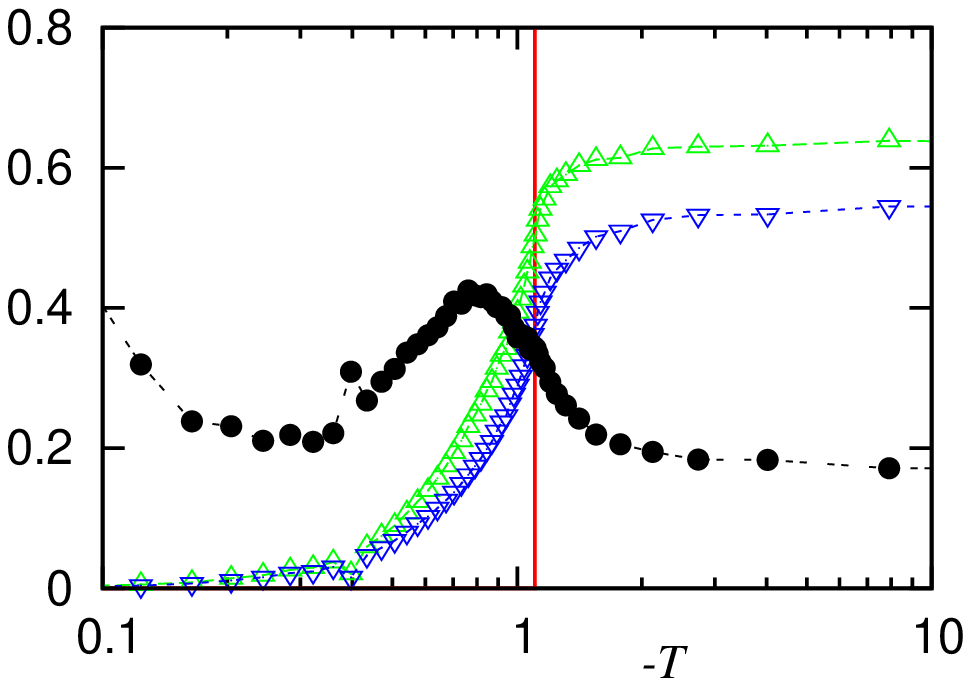,angle=270,width=6.3cm}
}
\caption{
\label{fig:heisaniso25}
Indicators of the phase transition and dynamic quantities as a function of temperature, for Heisenberg coupling and two types of anisotropic coupling.
For (a) and (b) $(J_x,J_y,J_z) = (1,1,1)/\sqrt{3} \approx (0.577,0.577,0.577)$, for (c) and (d) $(J_x,J_y,J_z) = (4,-2,1)/\sqrt{21} \approx (0.873, -0.436, 0.218)$, for (e) and (f) $(J_x,J_y,J_z) = (2,-1,1)/\sqrt{6} \approx (0.816, -0.408, 0.408)$, and for (g) and (h) $(J_x,J_y,J_z) = (1,1,0)/\sqrt{2} \approx (0.707, 0.707, 0.0)$.
The coupling constants are indicated above the plots.
Both the AF and the FM sides are shown.
Below some transitions, formation of magnetic domains prevents the averages and Lyapunov exponents from converging sufficiently within the simulation time.
}
\end{figure*}

Motivated by the above finding, we have systematically investigated the $G$-index for other interaction parameters.
Fig.~\ref{fig:heisaniso25} presents $\lm$, $\ks$, and $G$ at positive and negative temperatures for the Heisenberg interaction [(a) and (b)], generic anisotropic interaction [(c) and (d)], a less anisotropic interaction [(e) and (f)], and the $XX$-interaction [(g) and (h)]. The AF transitions in all these cases occurred at positive temperatures, while the FM transitions occurred at negative temperatures. Typically, as shown in Figs.~\ref{fig:heisaniso25}~(a)--(f) $G(T)$ exhibited a sharp peak at either FM or AF phase transitions but never at both. The $XX$-interaction was the only case in which we observed no peak in $G(T)$ at either of the two phase transitions. A further example illustrating the symmetry of the $G$-index with respect to the simulataneous sign change of the total energy and one of the coupling constants is given in Appendix~\ref{reversal}.

The behavior of $G(T)$ away from the phase transition, in particular the appearance of humps of $G(T)$ in Figs.~\ref{fig:heisaniso25}~(b), (g), and~(h), may also be of interest, but it extends beyond the scope of the present work. 
Here we only make two remarks: 
(i) This kind of humps should be distinguished from the ``peaks'' of the $G$-index associated with the phase transition. In the thermodynamic limit, the ``peaks'' are expected to have discontinuous first derivatives and hence be cusp-like. This is a consequence of the earlier general argument about the conversion from energy to temperature dependencies near the second-order phase transitions. On the contrary, the ``humps'' away from the second-order phase transitions are expected to remain broad and smooth in the thermodynamics limit. While the above distinction is  reasonably supported by our numerical results, the computational resources available to us were not sufficient to check the scaling of the $G$-index peaks near the phase transition beyond the results presented in Fig.~\ref{fig:finitesize-G} of Appendix~\ref{appendix:finitesize}.
(ii) Our calculations far into the ordered phases for generic anisotropic couplings of the type presented in Figs.~\ref{fig:heisaniso25}~(c), (d), (e), and (f) exhibited very slow convergence --- probably because of the formation of magnetic domains. In these two cases, we were not able to check whether humps similar to those seen in Figs.~\ref{fig:heisaniso25}~(b), (g), and~(h) exist at sufficiently low temperatures.

\section{Relation between the $G$-index and the Lyapunov vectors
\label{Discussion}}

Now we turn to explaining the presence or the absence of the peaks of $G(T)$ at $T=T_c$.
In general, all positive Lyapunov exponents tend to decrease with decreasing $|T|$, because the phase space volume available to the system becomes smaller.
The function $G(T)$ given by Eq.~(\ref{G}) is sensitive to the difference between the temperature dependences of $\lm$ and the average positive Lyapunov exponent $\lav \equiv \ks/N_s$.
Let us follow the behavior of $G(T)$ starting from infinite temperature and then decreasing $|T|$.
As can be seen in Figs.~\ref{fig:heisaniso25}~(a), (d), and~(f), $G(T)$ exhibits a peak at $T = T_c$ when 
$\lm$ initially decreases more slowly than $\lav$ and then drops faster around $T=T_c$, thereby catching up with $\lav$.
We now propose an argument, which we later substantiate by examples, that the above behavior of $\lm$ is due to the fact that the order parameter is capable of strong dynamical fluctuations.
In such a case, the Lyapunov vector corresponding to $\lm$ seeks the directions in the phase space corresponding to the faster-than-average dynamics, which are, in turn, correlated with the combinations of variables contributing to~$\phi$.
In the opposite case, when $\phi$ is not capable of sufficiently strong dynamical fluctuations, the Lyapunov vector corresponding to $\lm$ ignores the respective directions in the phase space.
In such a case, $\lm(T)$ and $\lav (T)$ exhibit very similar behavior over the entire range of temperatures seen in Figs.~\ref{fig:heisaniso25}~(b), (c), (e), (g) and~(h),  and, as a result, $G(T)$ does not have a peak at $T = T_c$.
  
In order to exemplify the notion of strong dynamical fluctuations of the order parameter, let us  assume that the magnetic order  sets in along the $x$-axis. [This is the only possibile direction for the interaction used for Figs.~\ref{fig:heisaniso25}~(c), (d), (e), and (f), or one of a continuous set of possible directions for Figs.~\ref{fig:heisaniso25}~(a), (b), (g) and~(h).] Let us then  decompose the Hamiltonian as 
\begin{eqnarray}
& {\mathcal H} = \sum_{m,n(m),m<n}  \bigg[ {1 \over 4} (J_y - J_z) (S_{m+} S_{n+} + S_{m-} S_{n-}) 
\nonumber
\\
&
+ {1 \over 4} (J_y + J_z) (S_{m+} S_{n-} + S_{m-} S_{n+})
 +J_x S_{mx} S_{nx} \bigg],
\label{H1}
\end{eqnarray}
where we use spin raising and lowering variables $ S_{m+} = S_{my} + i S_{mz}$ and $ S_{m-} = S_{my} - i S_{mz}$, which are analogous to the raising and lowering quantum spin operators~\cite{Fine-97,Fine-04}. We refer to the first two terms in the right-hand-side of Eq.~(\ref{H1}) as ``double-flip" and ``flip-flop" terms respectively. The first of them changes the $z$-projections of the two spins in the same direction, while the second one changes them in the opposite directions. The flip-flop term makes AF order fluctuate, while conserving the FM order. The double-flip term has the opposite effect.

For the Heisenberg Hamiltonian, $J_y = J_z$. Therefore, the double-flip term is zero, while the flip-flop term dominates.  As a result, the AF order strongly fluctuates in time near the phase transition, which, according to our argument, leads to the peak of $G(T)$ seen in Fig.~\ref{fig:heisaniso25}~(a). On the contrary, the FM order that sets in at negative temperatures does not fluctuate in time.  Accordingly, $G(T)$ does not exhibit a peak at $T=T_c$ in Fig.~\ref{fig:heisaniso25}~(b). 

For the Hamiltonian corresponding to Figs.~\ref{fig:heisaniso25}~(c) -- (f), $|J_y - J_z| > |J_y + J_z| $. Therefore, the double-flip term dominates. This leads to the peak of $G(T)$ at the FM transition and no peak at the AF transition.

For the XX-interaction corresponding to Figs.~\ref{fig:heisaniso25}~(e, d), $|J_y - J_z| = |J_y + J_z| $, i.e. the flip-flop and the double-flip terms have equal stength. This implies that the Lyapunov vector corresponding to $\lm$ does not particularly seek either FM or AF correlations.  As a result, there are no peaks of $G(T)$ at either FM or AF transition.

\begin{figure}
\epsfig{figure=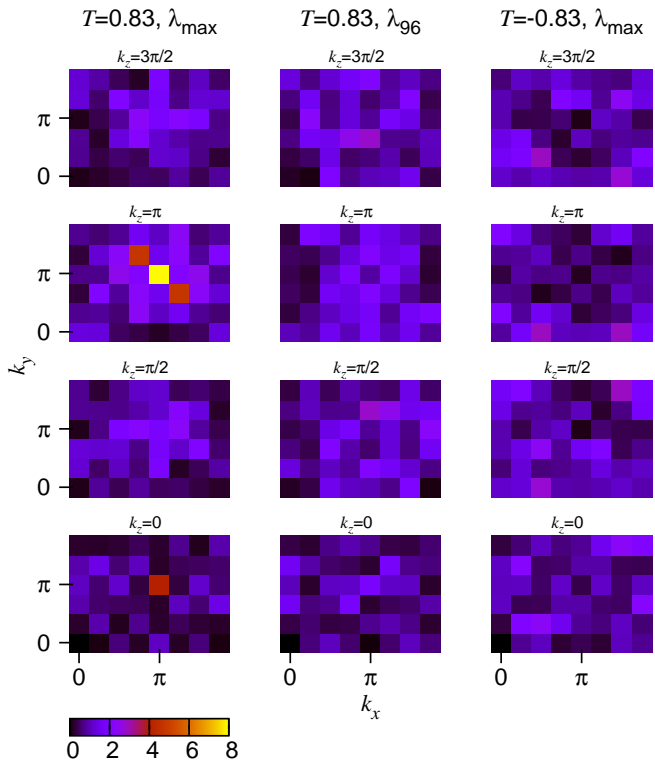,angle=270,width=8.6cm}
\caption{Spectral functions $ F(k_x, k_y, k_z) $ of three Lyapunov vectors for the lattice with Heisenberg interaction used for Figs.~\ref{fig:heisaniso25}~(a) and~(b). The left and the middle columns of frames represent $ F(k_x, k_y, k_z) $ for the Lyapunov vectors corresponding to $\lm$ and $\lambda_{96}$, respectively,  at the AF transition temperature. The right column corresponds to the Lyapunov vector for $\lm$ at the FM transition temperature.  Each column shows a complete set of values of $ F(k_x, k_y, k_z) $ encoded as color pixels: one pixel for each of  $8 \times 6 \times 4$ possible combinations of wave numbers $(k_x, k_y, k_z)$. Each frame contains $8 \times 6$ pixels for a fixed value of $k_z$ indicated above the frame.
\label{fig:powerspectrum}}
\end{figure}

The above interpretation is supported by our Fourier analysis of the components of Lyapunov vectors $\{  \delta \mathbf{S}_ {(n_x, n_y, n_z)} \}$~\cite{chaosspins2}.
Here $(n_x, n_y, n_z)$ are the cubic lattice indices.
We compute the function $ F(k_x, k_y, k_z) \equiv \mathbf{A}^*(k_x, k_y, k_z) \cdot \mathbf{A}(k_x, k_y, k_z)$, where
\begin{eqnarray}
\mathbf{A}(k_x, k_y, k_z) \equiv 
\sum_{n_x, n_y, n_z} \delta \mathbf{S}_ {(n_x, n_y, n_z)} e^{ -i (k_x n_x + k_y n_y + k_z n_z) }~,
\end{eqnarray}
and $(k_x, k_y, k_z)$ are the wave numbers of the discrete Fourier components.

$F(k_x, k_y, k_z)$ for the Heisenberg case is presented in Fig.~\ref{fig:powerspectrum}.
The left column of this figure shows $F(k_x, k_y, k_z)$ for the Lyapunov vector corresponding to $\lm$ at the temperature of the AF phase transition. For comparison, the middle column represents $F(k_x, k_y, k_z)$ for
the Lyapunov exponent $\lambda_{96}$ taken from the middle of the positive side of the Lyapunov spectrum at the same temperature and the right column corresponds to $\lm$  but at the (negative) temperature of the FM transition.
In the first case, the bright spots in Fig.~\ref{fig:powerspectrum} around $k_x=k_y=k_z = \pi$ indicate strong AF correlations. In the latter two cases, no correlations of AF or FM type are apparent.

\section{Possible generalisations\label{sec:alternativedynamics}}

The question arises as to how general are the results obtained in this article.
One possible generalization of the system considered here is a lattice of interacting rigid rotators.
In this case, the Hamiltonian would depend on the rotation angle and the angular momentum of each rotator.
The dynamics of such a system would include several aspects different from the classical spin lattices considered in this work.
Firstly, the dimensionality of the phase space and hence the number of the Lyapunov exponents would be two times larger.
Secondly, since the kinetic energy of the rigid rotator system is not limited from above, it cannot have negative temperature.
We do not expect the $G$-index of the rigid rotator lattices to be the same as for the classical spin lattices considered in this article.
At the same time, lattices of rigid rotators are also likely to exhibit a second-order phase transitions.
We would then expect that the $G$-index in this case, would also exhibit a peak at the phase transition for some but not for all possible interaction models.
Such an investigation, however, extends beyond the scope of the present article.
Another interesting potential investigation would be to compute the $G$-index for the
classical liquid-gas system near the critical point on the
pressure-temperature phase diagram, where the line of the first-order
phase transition ends.

\section{Conclusions}
\label{Conclusions}

To summarize, we have identified a characteristics of Lyapunov spectra of many-spin systems --- the $G$-index --- which, as a function of temperature exhibits a clear peak at magnetic phase transitions, provided the variable associated with the order parameter is capable of strong dynamic fluctuations. We expect similar behavior near second-order phase transitions in other many-particle systems with short-range interactions. As a part of this work, we have also developed an algorithm for determining microcanonical temperatures of general Hamiltonian systems.

\bigskip
{\it Note added:} Recently,
we discovered that a significant part of the justification of our temperature-determining algorithm
[namely, roughly that up to Eq.~(A6) in Appendix A] was done in Ref.~\cite{rugh}.

\acknowledgements{The authors are grateful to T. A. Elsayed for helpful discussions during the initial stage of this work.
A.S.dW's work is financially supported by an Unga Forskare grant from the Swedish Research Council.
The numerical part of this work was performed at the bwGRiD computing cluster at the University of Heidelberg.}

\appendix

\section{Derivation of Eq.(\ref{T-alg}) for temperature associated with a given energy shell}
\label{T-appendix}

Here we derive Eq.~(\ref{T-alg}) by perturbing an energy shell and making use Eq.~(\ref{T}).

Let us start by mentioning that an intuitive insight in the forthcoming general derivation can be gained by considering an example of $N$-dimensional Euclidean phase space, and assuming that the energy is given by the distance to the origin of a Cartesian coordinate system in this space.
In this case, the family of energy shells becomes a continuous set of $(N-1)$-dimensional hyperspherical surfaces with a common center.

Turning to the general case, let us denote the complete set of coordinates in $N$-dimensional many-particle phase space as $\{q_1, ..., q_N \} \equiv \vec{q}$.  An energy shell corresponding to energy $E_0$ is defined by condition
\begin{equation}
E(\vec{q}) = E_0.
\label{E0}
\end{equation}
The corresponding $(N-1)$-dimensional phase-space volume is $V(E_0)$. 
Let us further consider a small element of volume $\delta V(E_0)$ on this energy shell. If the energy changes by value $dE$, the above element can be bijectively mapped onto an element of the new energy shell by moving in the direction orthogonal to the original energy shell. The change of the coordinates in this case is 
\begin{equation}
\vec{q} \to \vec{q} + d \vec{q} ,
\label{qq}
\end{equation}
where
\begin{equation}
d \vec{q} = \vec{g} {dE \over | \vec{g}|^2} .
\label{dq}
\end{equation}
Here $\vec{g} \equiv \partial E(\vec{q})/\partial \vec{q}$ is the vector orthogonal to the original energy shell at a given point.  The individual components of this vector are  $g_i = \partial E(q_1, ..., q_N)/\partial q_i$.  The volume of the above element of the energy surface after transformation (\ref{qq}) is
\begin{equation}
\delta V(E_0 +d E) = \delta V(E_0) \ \det  \! \left[ 
\delta_{ij} + dE  {\partial \over  \partial q_i } \left( {g_j \over | \vec{g}|^2}  \right) 
\right],
\label{deltaV}
\end{equation}
where $\delta_{ij}$ is the Kronecker delta. Strictly speaking, the determinant in Eq.~(\ref{deltaV}) represents the growth of an $N$-dimensional rather than $(N-1)$-dimensional volume element, because it includes the energy direction itself. However, since this is only one of $N \gg 1$ directions, the error in the final result associated with the above approximation is of the order of $1/N$.
Now, we write explicitly
\begin{eqnarray}
\nonumber
&{\partial \over  \partial q_i } \left( {g_j \over | \vec{g}|^2}  \right) =  {1\over | \vec{g}|^2}  {\partial^2 E \over  \partial q_i  \partial q_j}  
\\
&\phantom{nnn} \strut- \frac{1}{| \vec{g}|^4}\frac{\partial E}{\partial q_j}  {{\partial \over  \partial q_i } \left[ \left( {\partial E \over \partial q_1}\right)^2 + ... + \left({\partial E\over \partial q_N}\right)^2  \right] } ,
\label{dgdq}
\end{eqnarray}
and then observe that for a system with short-range interactions, the first term in the above equation is of the order of $1/N$, while the second term is of the order of $1/N^2$ and hence can be neglected.  We further notice that the leading (first-order) contributions to the determinant in Eq.~(\ref{deltaV}) in terms of $dE$  come only from the diagonal elements of the matrix.  Taking into account the above two considerations, we finally obtain that, in the limit $N \to \infty$,
\begin{equation}
\delta V(E_0+d E) = \delta V(E_0) \left(1 + \frac{{\mathcal K}}{|\vec{g}|^2}  dE\right)~,
\end{equation}
where ${\mathcal K} = \sum_i \partial^2 E/\partial q_i^2 $.
The total change of the volume of the energy shell is then
\begin{equation}
V(E_0+dE) = {V(E_0)} \left( 1 + \left\langle \frac{{\mathcal K}}{|\vec{g}|^2} \right\rangle dE\right)~,
\label{deltaV1}
\end{equation}
where the notation $\langle ... \rangle$ implies the average of the entire energy shell. 

\begin{figure}
\epsfig{figure=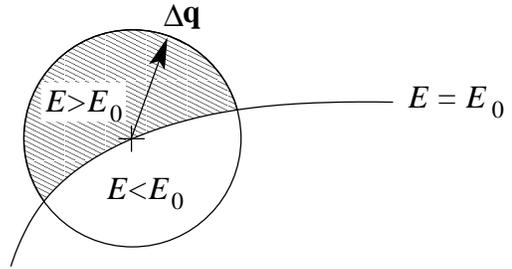,width=6.6cm}
\caption{A diagram illustrating how a point on an energy shell becomes displaced by a perturbation $d \vec{q}$ that changes the energy.  Due to the curvature of the energy shell, an increase in the energy is more likely than a decrease (for $T>0$).
\label{fig:energyshell}}
\end{figure}

Eq.~(\ref{deltaV1}) together with Eq.~(\ref{T}) implies that $1/T = \left\langle {\mathcal K} / |\vec{g}|^2 \right\rangle $.
Since both $|\vec{g}|^2$ and ${\mathcal K}$ contain additive small contributions associated with the uncorrelated remote parts of a large system, the distributions for each of them are narrowly peaked around the respective average values (according to the central limit theorem). Therefore, in the limit $N \to \infty$,
\begin{equation}
\frac{1}{T} =  \frac{\left\langle {\mathcal K}\right\rangle}{ \left\langle |\vec{g}|^2\right\rangle} ~.
\label{TKg}
\end{equation}

In our simulations, both $\left\langle {\mathcal K}\right\rangle$ and $\left\langle |\vec{g}|^2\right\rangle$ are obtained from  the perturbations of the phase space vector $\Delta \vec{q}$ associated with the small random spin rotations introduced in the main part of the paper and characterized by mean-squared values $\sigma_q^2 \equiv \left\langle \Delta q_i^2\right\rangle \ll 1/N $. These perturbations are illustrated schematically in Fig.~\ref{fig:energyshell}. The energy change for each perturbation is
\begin{equation}
\Delta E \approx \sum_i g_i  \Delta q_i + \frac{1}{2} \sum_{i,j} {\partial^2 E \over \partial q_i \partial q_j} \Delta q_i  \Delta q_j ~,
\end{equation}
from which it follows that, in the limit $\sigma_q^2 \to 0$, $\langle \Delta E \rangle = \frac12 \left\langle \mathcal K \right\rangle \sigma_q^2$, while $\langle \Delta E^2 \rangle = \left\langle |\vec{g}|^2\right\rangle \sigma_q^2$. Substituting the latter two formulas into Eq.~(\ref{TKg}), we obtain Eq.~(\ref{T-alg}). 

\begin{figure}
\epsfig{figure=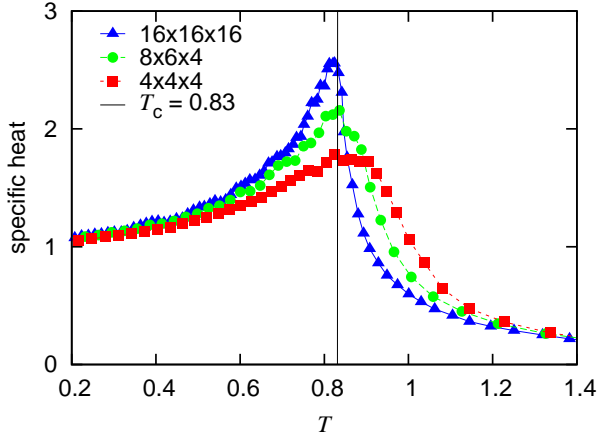,angle=270,width=8.6cm}
\caption{
Specific heat of the Heisenberg model for lattices of different sizes indicated in the figure.
\label{fig:C-Heisenberg}}
\end{figure}

\begin{figure}
\epsfig{figure=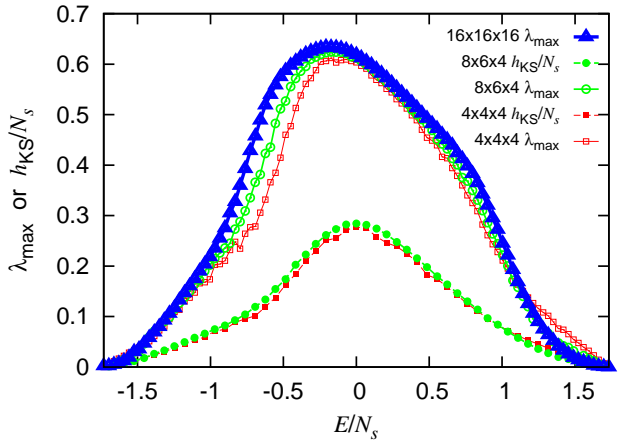,angle=270,width=8.6cm}
\caption{
The largest Lyapunov exponent and Kolmogorov-Sinai entropy per spin of the Heisenberg model for several different lattice sizes as functions of the energy per spin.
The Kolmogorov-Sinai entropy for the $16\times 16 \times 16$ system is not shown, as it is impossible to obtain sufficiently accurately with currently available computing power.
Even for the largest system, the Lyapunov exponents and Kolmogorov-Sinai entropy depend smoothly on the energy.
\label{fig:finitesizelyapunov-E}}
\end{figure}

\begin{figure}
\epsfig{figure=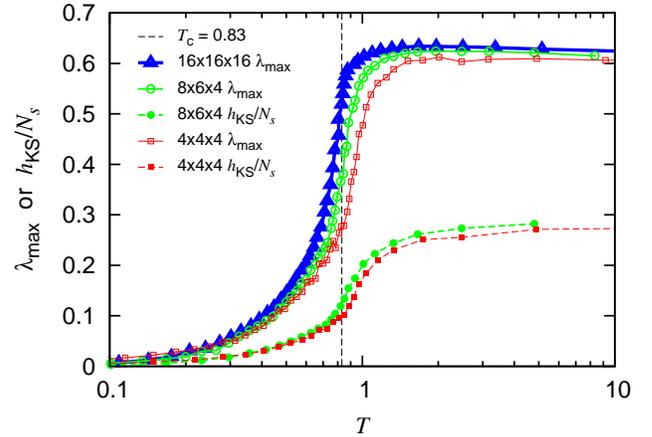,angle=270,width=8.6cm}
\caption{
The same quantities as in Fig.~\ref{fig:finitesizelyapunov-E} but plotted as functions of temperature, for $E<0$. 
\label{fig:finitesizelyapunov-T}}
\end{figure}

\begin{figure}
\epsfig{figure=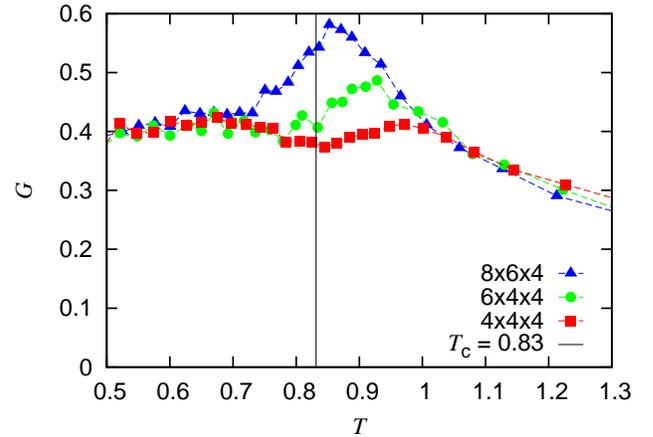,angle=270,width=8.6cm}
\caption{
Temperature dependences of the $G$-index for the Heisenberg model on lattices of three different sizes indicated in the figure.
\label{fig:finitesize-G}}
\end{figure}

\section{Finite-size effects\label{appendix:finitesize}}

Here we include four figures illustrating the dependence of several quantities computed for the Heisenberg model on lattice sizes:  Fig.~\ref{fig:C-Heisenberg} shows the size dependence of the specific heat, Figs.~\ref{fig:finitesizelyapunov-E} and \ref{fig:finitesizelyapunov-T} show the largest Lyapunov exponent and the Kolmogorov-Sinai entropy as functions of energy and temperature, respectively, and, finally, Fig.~\ref{fig:finitesize-G} shows the $G$-index.

\section{Simultaneous sign reversal of the total energy and one of the coupling constants}
\label{reversal}

\begin{figure*}
\parbox[t]{6.3cm}{\hskip9mm{\figfont (a)}\hfill\strut\\[-10mm]
    \epsfig{figure=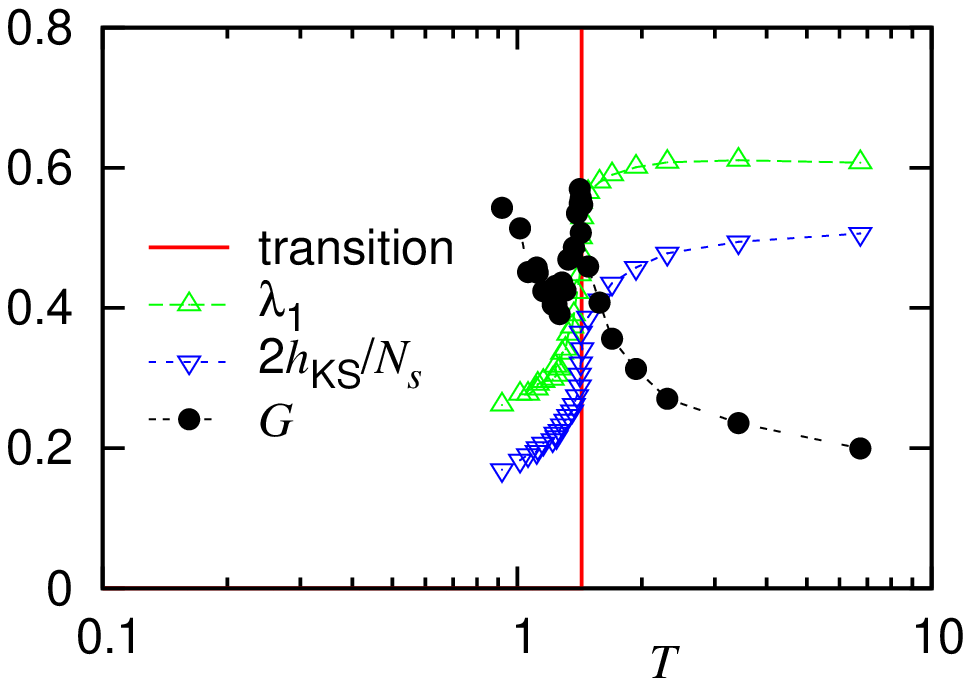,angle=270,width=6.3cm}
}
\parbox[t]{6.3cm}{\hskip9mm{\figfont (b)}\hfill\strut\\[-10mm]
    \epsfig{figure=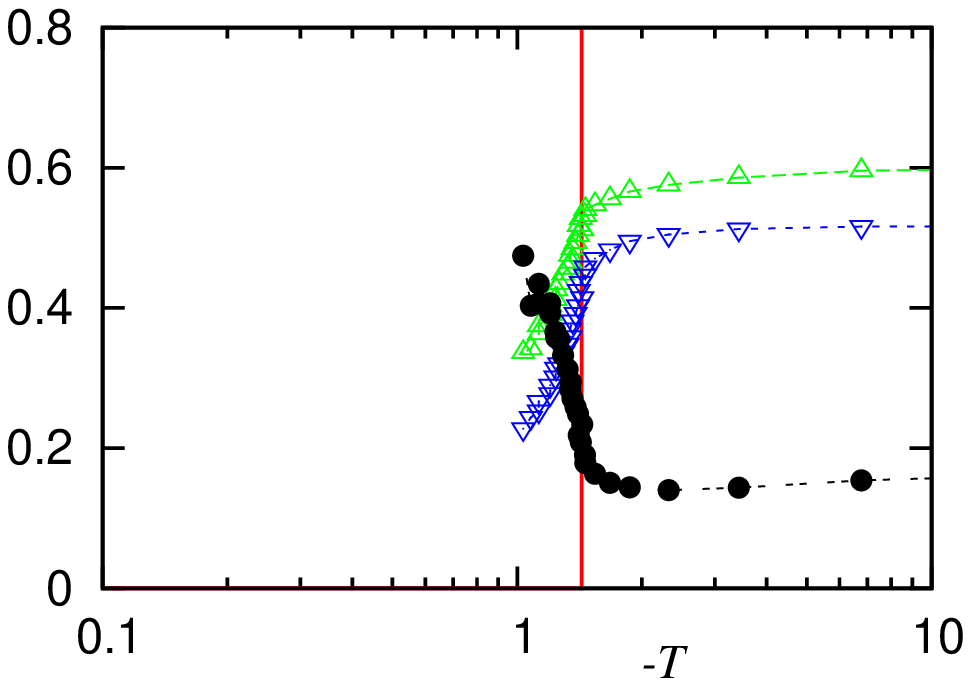,angle=270,width=6.3cm}
}\\[0.5ex]
\caption{\label{fig:collecteddatasupplemental}
Lyapunov exponents and $G$-index as functions of temperature for several different sets of coupling constants (indicated above the plots).  The position of the phase transition is indicated with a red vertical line.
$(J_x,J_y,J_z) = (4,2,1)/\sqrt{21} \approx (0.873, 0.436, 0.218)$.
}
\end{figure*}

Figure~\ref{fig:collecteddatasupplemental} shows plots similar to Figs.~\ref{fig:heisaniso25}~(c) and (d), but for a different sign of the couplings.
The comparison of Fig.~\ref{fig:collecteddatasupplemental}~(a) with Fig.~\ref{fig:heisaniso25}~(d), and Fig.~\ref{fig:collecteddatasupplemental}~(b) with Fig.~\ref{fig:heisaniso25}~(c) demonstrates the symmetry of Lyapunov spectra of bipartite spin lattices with respect to the simultaneous sign reversals of the total energy and of one of the coupling constants.

\end{document}